\PassOptionsToPackage{unicode}{hyperref}
\PassOptionsToPackage{hyphens}{url}
\PassOptionsToPackage{dvipsnames,svgnames,x11names}{xcolor}
\documentclass[
  12pt]{article}

\usepackage{amsmath,amssymb}
\usepackage{iftex}
\ifPDFTeX
  \usepackage[T1]{fontenc}
  \usepackage[utf8]{inputenc}
  \usepackage{textcomp} % provide euro and other symbols
\else % if luatex or xetex
  \usepackage{unicode-math}
  \defaultfontfeatures{Scale=MatchLowercase}
  \defaultfontfeatures[\rmfamily]{Ligatures=TeX,Scale=1}
\fi
\usepackage{lmodern}
\ifPDFTeX\else  
\fi
\IfFileExists{upquote.sty}{\usepackage{upquote}}{}
\IfFileExists{microtype.sty}{% use microtype if available
  \usepackage[]{microtype}
  \UseMicrotypeSet[protrusion]{basicmath} % disable protrusion for tt fonts
}{}
\makeatletter
\@ifundefined{KOMAClassName}{% if non-KOMA class
  \IfFileExists{parskip.sty}{%
    \usepackage{parskip}
  }{% else
    \setlength{\parindent}{0pt}
    \setlength{\parskip}{6pt plus 2pt minus 1pt}}
}{% if KOMA class
  \KOMAoptions{parskip=half}}
\makeatother
\usepackage{xcolor}
\makeatletter
\ifx\paragraph\undefined\else
  \let\oldparagraph\paragraph
  \renewcommand{\paragraph}{
    \@ifstar
      \xxxParagraphStar
      \xxxParagraphNoStar
  }
  \newcommand{\xxxParagraphStar}[1]{\oldparagraph*{#1}\mbox{}}
  \newcommand{\xxxParagraphNoStar}[1]{\oldparagraph{#1}\mbox{}}
\fi
\ifx\subparagraph\undefined\else
  \let\oldsubparagraph\subparagraph
  \renewcommand{\subparagraph}{
    \@ifstar
      \xxxSubParagraphStar
      \xxxSubParagraphNoStar
  }
  \newcommand{\xxxSubParagraphStar}[1]{\oldsubparagraph*{#1}\mbox{}}
  \newcommand{\xxxSubParagraphNoStar}[1]{\oldsubparagraph{#1}\mbox{}}
\fi
\makeatother

\usepackage{longtable,booktabs,array}
\usepackage{calc} % for calculating minipage widths
\usepackage{etoolbox}
\makeatletter
\patchcmd\longtable{\par}{\if@noskipsec\mbox{}\fi\par}{}{}
\makeatother
\IfFileExists{footnotehyper.sty}{\usepackage{footnotehyper}}{\usepackage{footnote}}
\makesavenoteenv{longtable}
\usepackage{graphicx}
\makeatletter
\def\maxwidth{\ifdim\Gin@nat@width>\linewidth\linewidth\else\Gin@nat@width\fi}
\def\maxheight{\ifdim\Gin@nat@height>\textheight\textheight\else\Gin@nat@height\fi}
\makeatother
\setkeys{Gin}{width=\maxwidth,height=\maxheight,keepaspectratio}
\makeatletter
\def\fps@figure{htbp}
\makeatother

\makeatletter
\@ifpackageloaded{caption}{}{\usepackage{caption}}
\AtBeginDocument{%
\ifdefined\contentsname
  \renewcommand*\contentsname{Table of contents}
\else
  \newcommand\contentsname{Table of contents}
\fi
\ifdefined\listfigurename
  \renewcommand*\listfigurename{List of Figures}
\else
  \newcommand\listfigurename{List of Figures}
\fi
\ifdefined\listtablename
  \renewcommand*\listtablename{List of Tables}
\else
  \newcommand\listtablename{List of Tables}
\fi
\ifdefined\figurename
  \renewcommand*\figurename{Figure}
\else
  \newcommand\figurename{Figure}
\fi
\ifdefined\tablename
  \renewcommand*\tablename{Table}
\else
  \newcommand\tablename{Table}
\fi
}
\@ifpackageloaded{float}{}{\usepackage{float}}
\floatstyle{ruled}
\@ifundefined{c@chapter}{\newfloat{codelisting}{h}{lop}}{\newfloat{codelisting}{h}{lop}[chapter]}
\floatname{codelisting}{Listing}

\makeatother
\makeatletter
\@ifpackageloaded{caption}{}{\usepackage{caption}}
\@ifpackageloaded{subcaption}{}{\usepackage{subcaption}}
\makeatother

\ifLuaTeX
  \usepackage{selnolig}  % disable illegal ligatures
\fi
\usepackage[]{natbib}
\usepackage{bookmark}

\IfFileExists{xurl.sty}{\usepackage{xurl}}{} % add URL line breaks if available
\hypersetup{
  pdftitle={Title},
  pdfauthor={Author 1; Author 2},
  pdfkeywords={3 to 6 keywords, that do not appear in the title},
  colorlinks=true,
  linkcolor={blue},
  filecolor={Maroon},
  citecolor={Blue},
  urlcolor={Blue},
  pdfcreator={LaTeX via pandoc}}

\newcommand{\anon}{1}

\usepackage{algorithm}
\usepackage{algorithmicx}
\usepackage{algpseudocode}
\usepackage{setspace}
\newtheorem{assumption}{Assumption}
\newtheorem{theorem}{Theorem}

\usepackage{xcolor}
\usepackage{amsmath}
\usepackage{amssymb}

\newcommand{\bA}{ \mbox{\bf A}}
\newcommand{\ba}{ \mbox{\bf a}}
\newcommand{\bB}{ \mbox{\bf B}}

\newcommand{\bx}{ \mbox{\bf x}}

\newcommand{\bZ}{ \mbox{\bf Z}}
\newcommand{\by}{ \mbox{\bf y}}

\newcommand{\bS}{ \mbox{\bf S}}

\newcommand{\bV}{ \mbox{\bf V}}

\newcommand{\bM}{ \mbox{\bf M}}

\newcommand{\bQ}{ \mbox{\bf Q}}
\newcommand{\bU}{ \mbox{\bf U}}

\newcommand{\argmin}{{\mathop{\rm arg\, min}}}

\newcommand{\beq}{ \begin{equation}}
\newcommand{\eeq}{ \end{equation}}
\newcommand{\beqn}{ \begin{eqnarray}}
\newcommand{\eeqn}{ \end{eqnarray}}
\newcommand{\bbb}{\boldsymbol}

\begin{document}

\def\spacingset#1{\renewcommand{\baselinestretch}%
{#1}\small\normalsize} \spacingset{1}

%%%%%%%%%%%%%%%%%%%%%%%%%%%%%%%%%%%%%%%%%%%%%%%%%%%%%%%%%%%%%%%%%%%%%%%%%%%%%%

\if1\anon
{
\title{\bf Spatial Causal Tensor Completion for Multiple Exposures and Outcomes: An Application to the Health Effects of PFAS Pollution}
\author{Xiaodan Zhou \\ 
Department of Statistics, North Carolina State University\\
Brian J Reich \\
Department of Statistics, North Carolina State University\\
Shu Yang\thanks{
The authors thank Dr. Ana Rappold of the United States Environmental Protection Agency for her help accessing and interpreting the data.  This work was partially supported by National Institutes of Health grants R01ES031651 and R01ES036270.} \\
Department of Statistics, North Carolina State University
% and \\ 
% Ana G Rappold\\
% Center for Public Health and Environmental Assessment, \\ US Environmental Protection Agency
}
  \maketitle
} \fi

\if0\anon
{
  \bigskip
  \bigskip
  \bigskip
  \begin{center}
    {\LARGE\bf Spatial Causal Tensor Completion for Multiple Exposures and Outcomes: An Application to the Health Effects of PFAS Pollution}
\end{center}
  \medskip
} \fi

\bigskip
\begin{abstract}
Per- and polyfluoroalkyl substances (PFAS) are typically encountered as mixtures of distinct chemicals with distinct effects on multiple health outcomes. Estimating joint causal effects using spatially-dependent observed data is challenging. We propose a spatial causal tensor completion framework that jointly models multiple exposures and outcomes within a low-rank tensor structure, while adjusting for observed confounders and latent spatial confounders. This method combines a low-rank tensor representation to pool information across exposures and outcomes with a spectral adjustment step that incorporates graph-Laplacian eigenvectors to approximate unmeasured spatial confounders, implemented via a projected-gradient descent algorithm. This framework enables causal inference in the presence of unmeasured spatial confounding and pervasive missingness of potential outcomes. We establish theoretical guarantees for the estimator and evaluate its finite-sample performance through extensive simulations. In an application to national PFAS monitoring data, our approach yields more conservative and credible causal relationships between PFOA and PFOS exposure and 13 chronic disease outcomes compared with existing alternatives. 
\end{abstract}

\noindent%
\vfill

\newpage
\spacingset{1.8} % DON'T change the spacing!

\section{Introduction}\label{s:intro}

Per- and polyfluoroalkyl substances (PFAS) are a class of synthetic chemicals widely used since the 1950s due to their resistance to water, oil, and heat \citep{gaines2023historical}. Because of their persistence in the environment, they are often referred to as `forever chemicals,' raising substantial public health concerns. PFAS chemicals rarely occur in isolation; drinking water contamination typically involves mixtures of compounds such as PFOA and PFOS. Multiple PFAS chemicals have been associated with a range of adverse outcomes, including altered immune and thyroid function, liver disease, lipid and insulin dysregulation, kidney disease, adverse reproductive and developmental outcomes, and cancer \citep{fenton2021per}. Yet despite growing attention, evidence for many health outcomes remains inconsistent and inconclusive \citep{national2022guidance}, potentially due to the complexity of the problem.

In this work, we study the effects of PFAS exposures on multiple health outcomes. This naturally gives rise to a multiple-exposure, multiple-outcome setting. The analysis is complicated by two major challenges: first, PFAS are typically encountered as mixtures of distinct chemicals with different sources and health effects, making it difficult to isolate the contributions of individual components; second, both exposures and outcomes exhibit spatial patterns, introducing the possibility of unmeasured spatial confounding \citep{reich2021review}. These complexities motivate the development of new statistical approaches tailored to this problem.

A large statistical literature has developed to quantify the effects of exposure mixtures \citep{joubert2022powering, kang2023partial}. Widely used approaches include weighted quantile sum regression \citep{carrico2015characterization}, Bayesian kernel machine regression \citep{bobb2015bayesian}, and quantile-based g-computation \citep{keil2020quantile}. However, they generally focus on a single outcome and do not account for unmeasured spatial confounding. Univariate-outcome spatial methods have been proposed to adjust for spatially structured confounders \citep{thaden2018structural, papadogeorgou2019adjusting, osama2019inferring, keller2020selecting, schnell2020mitigating, marques2022mitigating, dupont2022spatial+, guan2023spectral, gilbert2021causal, wiecha2025two} but these approaches are limited to one exposure and one outcome at a time, making them unsuitable for analyzing PFAS mixtures with multiple health endpoints.

% While there is an extensive literature on statistical methods for quantifying the effects of exposure mixtures \citep{joubert2022powering, kang2023partial}, most of these approaches do not explicitly incorporate causal inference techniques, e.g., \cite{carrico2015characterization}, \cite{bobb2015bayesian}, \cite{hao2018model}, \cite{ferrari2020identifying}, \cite{antonelli2020estimating}, \cite{wei2020sparse}, \cite{roy2021perturbed}, \cite{gibson2021bayesian}, \cite{kowal2021bayesian}, \cite{chen2022statistical}. \cite{rosato2022investigate} review the PFAS literature and conclude that weighted quantile sum \cite{carrico2015characterization} and Bayesian kernel machine \cite{bobb2015bayesian} regressions are most commonly used to study PFAS mixtures. 
% However, these non-spatial methods fail to account for spatial dependence or missing spatial confounders, such as a social vulnerability index, that display spatial patterns and are correlated with both exposures and outcomes. Univariate methods have been proposed to adjust for missing spatial confounders (\cite{thaden2018structural}, , \cite{papadogeorgou2019adjusting}, \cite{osama2019inferring}, \cite{keller2020selecting}, \cite{schnell2020mitigating}, \cite{marques2022mitigating}, \cite{dupont2022spatial+}, \cite{guan2023spectral}, \cite{gilbert2021causal}). In summary, there is no unified statistical framework that can be applied to the proposed PFAS data analyses that account for spatial confounding in a multivariate analysis with a mixture of exposure variables. 

Parallel to these developments, existing work has reframed causal effect estimation as a matrix or tensor completion problem, where counterfactual outcomes are recovered under low-rank structures \citep{poulos2021retrospective, auerbach2022tensor, zhen2024nonnegative, mao2024mixed}. This line of research enables principled imputation of missing potential outcomes and extends naturally to multivariate settings. However, these methods typically treat each exposure independently, overlooking latent relationships among exposures. More recent approaches integrate exposures into a single tensor representation, allowing for joint modeling of exposure effects and improved efficiency, as demonstrated in \citet{abadie2024doubly}, \citet{agarwal2020synthetic}, \citet{gao2024causal}, and \citet{gao2025causal}.

Two recent parallel works also address multiple exposures and outcomes with spatial confounding, though with different strategies. \citet{prim2025spectral} propose a spectral confounder adjustment that uses a three-way tensor over exposure, outcome, and spatial scale to model causal effect,  
% relying on canonical polyadic tensor decomposition and 
assuming that confounding bias dissipates at more local spatial scales. Our approach is more general in that we use a tensor for the potential outcomes, rather than using tensors as components of a linear model. \citet{wu2025latent} develop a latent factor panel approach for spatiotemporal data that leverages repeated temporal observations to identify causal effects under a factor confounding assumption, where unmeasured confounders are captured by low-dimensional latent factors that evolve over time, combined with partial interference assumptions that limit the degree of spatial spillover. By contrast, we propose a spatial causal tensor completion framework specifically designed for cross-sectional spatial data with multiple binary exposures and multiple outcomes. 

Our method combines a low-rank Tucker tensor representation \citep{kolda2009tensor} to pool information across exposures and outcomes with a spectral adjustment step that incorporates graph-Laplacian eigenvectors to approximate unmeasured spatial confounders, implemented via a spatial projected-gradient descent algorithm. Unlike panel approaches that exploit temporal variation, our framework relies on spatial smoothness assumptions to enable causal inference in the presence of spatially patterned unmeasured confounding and pervasive missingness of potential outcomes. We establish rigorous theoretical guarantees, including Frobenius error bounds for the estimated potential outcomes tensor that decompose into interpretable components, and provide uncertainty quantification of average treatment effects through doubly robust variance estimation. We demonstrate the framework's advantages through extensive simulations and an application to an national PFAS monitoring data. Our analysis covers 5,495 public water systems across 47 states and 13 chronic disease outcomes, reflecting typical contemporary drinking-water exposure levels rather than extreme contamination scenarios. We find that most previously reported PFAS–health associations are substantially attenuated after adjustment for spatial confounding. However, PFOS exposure remains significantly associated with elevated risk of hypertension, tooth loss, asthma, and obesity—associations that persist after adjustment and thus appear less likely to reflect geographic artifacts alone. These findings underscore the importance of accounting for spatial confounding in environmental health studies and suggest that targeted concerns about PFOS warrant continued attention.

This work makes three primary contributions. First, we develop a unified causal inference framework that jointly models multiple exposures and outcomes through low-rank tensor completion, enabling efficient estimation of factorial treatment effects in environmental mixtures studies. Second, we integrate graph-Laplacian spectral adjustment into the tensor completion procedure to account for unmeasured spatial confounding—a pervasive challenge in environmental epidemiology. Third, we establish theoretical guarantees for the proposed estimator and validate its finite-sample performance through simulations. Our application to national PFAS data provides a large-scale, spatially-adjusted assessment of drinking-water PFOA and PFOS exposure on chronic disease outcomes, yielding more conservative and credible effect estimates than existing methods.

The remainder of the paper is organized as follows. Section \ref{s:data} describes the motivating dataset. Sections \ref{s:method} and \ref{s:theory} present the proposed methodology and its theoretical properties, with additional empirical evaluations reported in \textbf{Supplement D}. Section \ref{s:app} applies the proposed approach to the motivating data. Section \ref{s:discussion} concludes.

\section{PFAS Data Description}\label{s:data}

% \subsection{Data Description}\label{s:data_desc}

The EPA’s PFAS Analytics Tools \citep{epapfastool} provide national-level PFAS data in Public Water Systems (PWS). UCMR5 reports multiple PFAS chemicals. Among these, PFOA and PFOS are the most frequently detected compounds, while the remaining chemicals were detected at only a small number of locations. Therefore, we focus on PFOA and PFOS in this study.

The EPA establishes minimum reportable limits (MRL) as the lowest measurable concentration of a contaminant that is achievable by the majority of participating laboratories, independently of the contaminant health effects. For this reason, most often only the presence or absence of the chemical above the MRL is used as the exposure measure. In addition, the health-based Maximum Contaminant Level Goals (MCLGs) defined by EPA for PFOA and PFOS are both zero ppt (also expressed as ng/L). Therefore, the exposure is taken to be the binary indicator of detectable PFOA and/or PFOS, given four possible exposure levels: Neither PFOA or PFAS, only PFOA exposure, only PFOS exposure, and both PFOA and PFOS exposures.

We use health outcome data provided by the Centers for Disease Control and Prevention (CDC) PLACES dataset \citep{centers2023places}, which contains model-based census tract-level estimates of disease prevalence among the adult population. The 13 health outcomes are: arthritis (ARTHRITIS), hypertension (BPHIGH), cancer (non-skin) or melanoma (CANCER), asthma (CASTHMA), coronary heart disease (CHD), chronic obstructive pulmonary disease (COPD), diagnosed diabetes (DIABETES), high cholesterol among adults who have ever been screened (HIGHCHOL), chronic kidney disease (KIDNEY), depression (MHLTH), obesity (OBESITY), stroke (STROKE), and all tooth loss among adults aged $\geq$65 years (TEETH). Demographic covariates were obtained from the 2020 US Census Bureau via the \texttt{tidycensus} R package \citep{walker2021package}; specific covariates are described in Section~\ref{s:app}.

Health outcomes and demographic covariates were linked to PWS using area-weighted averaging to aggregate tract-level variables to the PWS level (\textbf{Supplement C.1}). This yielded 5,495 PWS samples across 47 states in the continental United States. The Table in \textbf{Supplement C.2} % \ref{t:data_summary} 
summarizes PWS-level exposure groups and outcomes. 

% \subsection{Data Processing}\label{s:data_proc}

\section{Spatial Causal Tensor Model}\label{s:method} 

In this section, we introduce our spatial causal tensor framework and estimation strategy. Section \ref{s:notation} introduces notations, Section \ref{s:tensorframework} formalizes the potential-outcomes setup, and Section \ref{s:causal_assumption} states the required causal assumptions. Section \ref{s:integrated} then outlines a three-step estimator with a spatial projected gradient descent (S-PGD) algorithm. 

\subsection{Notations and Tensor Operations}
\label{s:notation}

Lowercase letters (e.g., $x, y$) denote scalars; bold lowercase letters (e.g., $\bx, \by$) denote vectors; bold uppercase letters (e.g., $\bA, \bB$) denote matrices; calligraphic uppercase letters (e.g., $\mathcal{X}, \mathcal{Y}, \mathcal{Z}$) denote tensors. A tensor $\mathcal{X} \in \mathbb{R}^{N_1 \times N_2 \times N_3}$ is a 3-way array indexed by $(i_1, i_2, i_3)$, where $i_k \in [N_k]$, and $[N_k] = \{ 1, ..., N_k\}$. The Frobenius norm of $\mathcal{X}$ is defined as $\|\mathcal{X}\|_F^2 = \sum_{i_1, i_2, i_3} \mathcal{X}_{i_1\, i_2\, i_3}^2$. The spectral and nuclear norms of tensors are defined formally in \textbf{Supplement A}. 
% \textcolor{red}{ Definitions~\ref{def:tensor_spectral} and \ref{def:tensor_nuclear}
The inner product is $\langle \mathcal{X}, \mathcal{Y}\rangle = \sum_{i,l,o} \mathcal{X}_{i\ell o} \mathcal{Y}_{i\ell o}$.  
The mode-\( k \) unfolding of a tensor \( \mathcal{X} \in \mathbb{R}^{N_1 \times \cdots \times N_d} \) rearranges the tensor into a matrix by aligning the mode-\( k \) fibers, a one-dimensional slice of a tensor obtained by fixing all indices but one, as columns of the resulting matrix, 
$
\mathcal{X}_{(1)} \in \mathbb{R}^{N_1 \times (N_2 N_3)},
\mathcal{X}_{(2)} \in \mathbb{R}^{N_2 \times (N_1 N_3)},
\mathcal{X}_{(3)} \in \mathbb{R}^{N_3 \times (N_1 N_2)}.
$
The mode-$k$ tensor-matrix product, denoted $\mathcal{X} \times_k \bM$, with $\bM \in \mathbb{R}^{r \times N_k}$, yields a tensor of dimension where mode $k$ is replaced by $r$. For example, $\mathcal{X} \times_1 \bM \in \mathbb{R}^{r \times N_2 \times N_3}$. This operation involves unfolding $\mathcal{X}$ along mode $k$, applying matrix multiplication, and refolding to the tensor form. For a comprehensive introduction to tensor operations, see \cite{kolda2009tensor}. 

% \textcolor{red}{
% Related inequalities are provided in Lemmas~\ref{lem:tensor_duality}--\ref{lem:tensor_nuclear_frob}, and a Bernstein-type concentration inequality for tensors is stated in Lemma~\ref{lem:tensor_Bernstein}.
% }

\subsection{Tensor Formulation of Potential Outcomes}\label{s:tensorframework}

Our proposed method is a novel attempt to impute the missing potential outcomes using low-rank tensor completion under the potential outcomes framework. Suppose there are \(N\) observational units, each exposed to a \(K\)-dimensional binary exposure vector \(\bA_i \in \{0,1\}^K\), which defines \(L = 2^K\) possible exposure combinations. We index each exposure combination by its decimal equivalent, denoted $\ell_i \in \{1,\dots,L\}$. For clarity, let $\mathbf{a}^{(\ell)} \in \{0,1\}^K$ represent the binary exposure vector corresponding to exposure level $\ell$. For example, when $K=2$, if $\bA_i = (A_{i1}, A_{i2})$, then $\ell_i = A_{i1} + 2A_{i2}$. In the motivating example, N = 5,495 denotes the number of geographic units (PWS) and K = 2 represents two PFAS exposure variables (PFOA and PFOS).

\begin{figure}[t]
\centering
\begin{minipage}{0.5\textwidth}
\centering
\includegraphics[width=\linewidth]{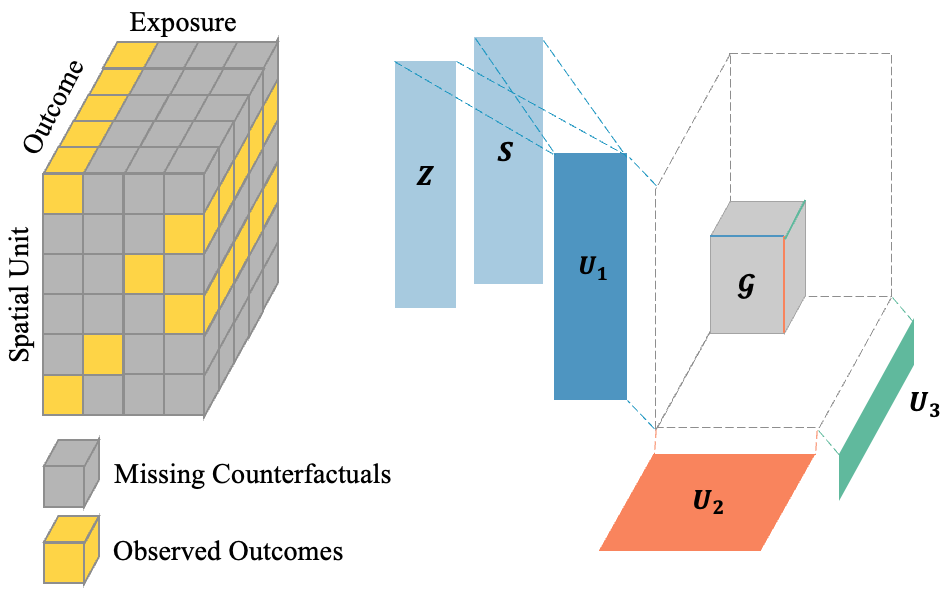}
\end{minipage}\hfill
\begin{minipage}{0.45\textwidth}
\caption{Illustration of the spatial causal tensor decomposition. Left: a tensor of counterfactual outcomes, where observed entries (yellow) and missing entries (gray) are arranged by spatial units (PWS), exposures, and outcomes. Right: Tucker decomposition into spatial ($\bU_1$), exposure ($\bU_2$), and outcome ($\bU_3$) factors with a low-rank core $\mathcal{G}$. The spatial factor $\bU_1$ is further expressed as a combination of measured covariates $\bZ$ and latent spatial components $\bS$. The illustration omits the noise term $\mathcal{E}$ for clarity.} 
\label{fig:illu}
\end{minipage}
\end{figure}

Under the potential outcomes framework \citep{rubin1974estimating}, each unit $i$ has a set of potential outcomes $\{\mathcal{Y}_{ilo}: \ell = 1,\dots,L,\; o=1,\dots,O\}$, one for every possible exposure combination $\ell$ and outcome dimension $o$. In the PFAS analysis, O = 13 outcomes are considered, as listed in Section \ref{s:data}. Only the outcome corresponding to the realized exposure is observed, while the others remain counterfactual. We organize these outcomes into a tensor $\mathcal{Y} \in \mathbb{R}^{N \times L \times O}$ (left of Figure~\ref{fig:illu}). Similarly, the treatment assignment is represented by the tensor \(\mathcal{A} \in \mathbb{R}^{N \times L \times O}\), with \( \mathcal{A}_{i\ell o} = 1 \) if unit \( i \) received exposure \( \ell \), and \( \mathcal{A}_{i\ell o} = 0 \) otherwise. Note that $\mathcal{A}_{i\ell o}$ does not vary across the third outcome-mode dimension, so it effectively reduces to an $N\times L$ matrix. We maintain the tensor representation for consistency with the outcome tensor and to simplify presentation. Each unit receives one exposure, i.e. $\sum_{\ell} \mathcal{A}_{i\ell o} = 1$. 

We assume that the observed outcome tensor follows a low-rank Tucker decomposition (Right of Figure~\ref{fig:illu}):
\begin{align}
&\mathcal{Y} = \mathcal{Y}^* + \mathcal{E} = \mathcal{G} \times_1 \bU_1\times_2 \bU_2 \times_3 \bU_3 + \mathcal{E}, \quad \mathcal{E} \stackrel{iid}{\sim} \text{sub-Gaussian}(0,\sigma^2). \label{eq:main}
\end{align}
In (\ref{eq:main}), \(\mathcal{G} \in \mathbb{R}^{r_1 \times r_2 \times r_3}\) is the core tensor and \(\bU_1 \in \mathbb{R}^{N \times r_1}\), \(\bU_2 \in \mathbb{R}^{L \times r_2}\) and \(\bU_3 \in \mathbb{R}^{O \times r_3}\) are factor matrices corresponding to spatial, exposure, and outcome variation, respectively. To capture both measured and unmeasured sources of spatial variation, we model $\bU_1 = \bZ\,\bbb\eta_Z + \bS\,\bbb\eta_S$, where \(\bZ\in\mathbb{R}^{N\times r_1}\) are measured covariates with effect \(\bbb\eta_Z\in\mathbb{R}^{r_1\times r_1}\), 
and \(\bS\in\mathbb{R}^{N\times r_1}\) are unmeasured spatial components with effect \(\bbb\eta_S\in\mathbb{R}^{r_1\times r_1}\). By unmeasured spatial confounder, we refer to latent variables that exhibit spatial structure and simultaneously influence exposures and outcomes. The measured covariates $\bZ$ and unmeasured spatial confounding $\bS$ jointly affect both the outcomes and the treatment assignment mechanism, as described below.  

%\begin{figure}%[tbp]
%\centering
%\includegraphics[width=.56\linewidth]{figs2/illustration2.png} 
%\caption{Illustration of the spatial causal tensor decomposition. Left: a tensor of counterfactual outcomes, where observed entries (yellow) and missing entries (gray) are arranged by spatial units (PWS), exposures, and outcomes. Right: Tucker decomposition into spatial ($\bU_1$), exposure ($\bU_2$), and outcome ($\bU_3$) factors with a low-rank core $\mathcal{G}$. The spatial factor $\bU_1$ is further expressed as a combination of measured covariates $\bZ$ and latent spatial components $\bS$. The illustration omits the noise term $\mathcal{E}$ for clarity.} 
%\label{fig:illu}
%\end{figure}

In this representation, $r_1$, $r_2$ and $r_3$ are less than or equal to $N$, $L$, and $O$, respectively. The factor matrices $\bU$ can be viewed as the principal components in each mode, and the entries in $\mathcal{G}$ entail the level of interaction between different components. The low-rank assumption on $\mathcal{Y}^*$ can be justified in the following way. First, the heterogeneity among locations may be attributed to $r_1$ latent spatial factors; e.g., sociodemographic trends. Second, the variation from the PFAS exposure may be summarized into $r_2$ key factors, such as chemicals with shared sources or physical, cheminformatic properties. Finally, the multiple health responses may be related due to $r_3$ hidden biological pathways. 

Our parameter of interest is the average treatment effect (ATE) defined for any exposure \(\ell\) (relative to the reference level \(L\)) and outcome \(o\) by:
$\theta_{\ell o}^*=\mathbb{E}\big(\sum_{i=1}^{N} \left(\mathcal{Y}_{i\ell o}-\mathcal{Y}_{iLo}\right)/N\big).$

\subsection{Causal Assumptions}\label{s:causal_assumption}

To formalize the exposure mechanism, the exposure probability for unit $i$ (with covariates $\mathbf{Z}_i$ and unobserved spatial factor $\bS_i$) to receive exposure $\ell$ is
\begin{align}
&\pi(\ell\mid \bZ_i, \bS_i) = \mathbb{P}(\bA_i = \mathbf{a}^{(\ell)} \mid \bZ_i, \bS_i).
\end{align}
We adopt the following classical causal assumptions to justify our estimating strategy.

\begin{assumption} [SUTVA; Stable Unit Treatment Value Assumption]
\label{sutva}
(1) the potential outcomes for any unit do not vary with the treatment assigned to other units; 
(2) there are no different versions of each treatment level leading to different potential outcomes.
\end{assumption}

Assumption~\ref{sutva} rules out interference between geographic units and requires that each exposure condition is well-defined. The no-interference component requires that PFAS exposure in one geographic unit does not affect health outcomes in another. This is plausible in our setting because exposure is measured through drinking water systems that serve distinct populations. Potential violations could nonetheless arise through population mobility or environmental contamination spreading between neighboring regions. We interpret our results as direct effects of local drinking water exposure, acknowledging that spillover effects, if present, are not captured. The consistency component requires careful interpretation: our binary exposure indicator aggregates heterogeneous concentration levels above the detection limit. Consequently, estimated effects represent averages over the observed distribution of exposure intensities rather than effects of a precisely defined exposure level.

\begin{assumption} [Latent Ignorability] %No unmeasured confounding
\label{nuc}
$ \bA_i \perp \{\mathcal{Y}_{i\ell o}: \ell,o\} \mid (\bZ_i,\bS_i)$. In other words, $\bZ_i$ and $\bS_i$ account for all confounders influencing treatments and outcomes. 
\end{assumption}

Assumption~\ref{nuc} extends the standard no-unmeasured-confounding condition by allowing for unobserved spatial confounders, such as unmeasured industrial activity or correlated environmental exposures that vary smoothly over space. We model these through low-dimensional latent spatial components approximated via spectral adjustment using the graph Laplacian eigenbasis. This approach is effective for confounders varying at spatial scales captured by the selected eigenvectors, but may fail for confounders varying at finer scales than our adjacency structure can resolve or for confounders that lack spatial structure altogether. The main threats to this assumption include: non-spatial confounders not captured by our measured covariates (such as bottled water); and temporal misalignment, as our exposure measurements (2023--2024) and health outcomes (2017--2018) span different periods, though the long environmental persistence and biological half-lives of PFAS partially mitigate this concern.

\begin{assumption}[Latent Positivity]\label{positivity}
There exists a constant $p_{\min}\in(0,\tfrac12)$ such that for every unit $i$ and exposure level $\ell\in\{1,\dots,L\}$,
\(
p_{\min}\ \le\ \mathbb{P}\,\!\big(\bA_i=\ba^{(\ell)}\mid \bZ_i,\bS_i\big)\ \le\ 1-p_{\min}.
\)
That is, each unit has a nonzero probability of receiving any exposure combination, ruling out deterministic assignment by covariates or location.
\end{assumption}

Assumption~\ref{positivity} requires that each exposure combination has nonzero probability of occurring, ruling out deterministic assignment by geography or covariates. Limited overlap is a known challenge in environmental studies, particularly when contamination is highly localized. We empirically assess this assumption by examining the distribution of estimated propensity scores in our application (\textbf{Table 3 in Supplement C.2}). The results indicate moderate overlap: only limited units have propensity scores below 0.01, though a notable fraction of exposed units (10 to 48\%) have scores below 0.05, with the PFOA-only and PFOS-only groups exhibiting the most limited support. 
% Our simulation study explicitly examines scenarios with weak overlap to assess robustness under such conditions.

\subsection{A Three-Step Estimating Procedure}\label{s:integrated} 

% Our estimator addresses three coupled tasks: estimate the latent spatial structure $\widehat{\bS}$ (to mitigate unmeasured spatial confounding); estimate exposure probabilities and weights $\widehat{\mathcal{W}}$ (to correct exposure-induced missingness); estimate the full potential-outcome tensor $\widehat{\mathcal{Y}}$ (to recover counterfactuals and ATEs). A standard PGD algorithm can recover low-rank tensor structure given a projection subspace, but cannot by itself estimate the exposure probabilities. We therefore adopt a three-stage approach in which a S-PGD simultaneously learns the tensor factors and the spatial component $\bS$, while a separate model estimates the propensity scores. 

We proceed in three steps. Step 1 uses an unweighted S-PGD fit with a Laplacian eigenbasis to learn the latent spatial component $\widehat{\bS}^{(0)}$ while estimating the Tucker factors from observed entries. Step 2 fits the exposure model $\widehat{\pi}(\ell\mid\bZ_i,\widehat{\bS}^{(0)}_i)$ and constructs inverse-probability weights $\widehat{\mathcal{W}}^{(0)}$ that account for exposure-induced missingness. Step 3 re-runs S-PGD with the weights to re-estimate the spatial component and complete the full potential-outcome tensor, yielding $\widehat{\bS}$ and $\widehat{\mathcal{Y}}$, and thus the final ATE estimates.

\subsubsection{Step 1: Spectral adjustment for spatial confounding}

Let $\bQ$ be the graph Laplacian built from the spatial adjacency matrix, with eigendecomposition $\bQ=\bbb\Phi\,\bbb\Lambda\,\bbb\Phi^\top$. Low-frequency eigenvectors in $\bbb\Phi$ capture broad spatial trends. We assume the unmeasured spatial component $\bS$ follows a Gaussian Markov random field that can be well-approximated by the span of the first $k$ low-frequency eigenvectors,
$$
\bS\approx \bbb\Phi_{1:k}\,\bbb\beta,\qquad \bU_1\approx\bZ\,\bbb\eta_Z+\bbb\Phi_{1:k}\,\bbb\beta,
$$
we denote $[\bbb\Phi_1, \bbb\Phi_2, ..., \bbb\Phi_k]$ as $\bbb\Phi_{1:k}$, with $\bbb\beta \in \mathbb{R}^{k\times r_1}$ are the corresponding coefficients. We obtain \(\widehat{\mathbf{S}}^{(0)}\) by minimizing the unweighted reconstruction loss  
\[
\argmin_{\,\mathcal{G}, \textbf{U}_2, \textbf{U}_3, \bbb\eta_Z, \bbb\beta} \bigg\{ || \mathcal{A} \circ (\mathcal{Y}-\mathcal{G}\times_1 (\bZ\,\bbb\eta_Z + \bbb\Phi_{1:k}\,\bbb\beta) \times_2 \bU_2 \times_3 \bU_3)||^2_F \bigg\}.
\]
using the S-PGD solver described in Section~\ref{s:comp_pgd}. This step provides an initial estimate of the latent spatial component. 
% \textcolor{red}{ I tried to soft the sentence here to avoid the impression that the whole algorithm relies on this step.}

\subsubsection{Step 2: Propensity score and weight estimation}

By Assumptions~(\ref{sutva})--(\ref{positivity}), inverse-probability weighting creates a pseudo-population in which exposure assignment is independent of the potential outcomes. Consequently, the exposure-induced missingness in $\mathcal{Y}$ is ignorable after weighting, enabling unbiased tensor completion under the Tucker model. Specifically, we estimate the exposure probabilities \(\widehat{\pi}(\ell \mid \bZ_i, \widehat{\mathbf{S}}^{(0)}_i)\) using multinomial logistic regression, and further construct inverse probability weights $\widehat{\mathcal{W}}^{(0)}$, 
\[
\widehat{\pi}(\ell | \bZ_i, \widehat{\mathbf{S}}^{(0)}_i) = \mathbb{P}(\bA_i = \mathbf{a}^{(\ell)} \mid \bZ_i, \widehat{\mathbf{S}}^{(0)}_i), \quad 
\widehat{\mathcal{W}}^{(0)}_{i\ell o} = \frac{\mathbf{1}(\bA_i = \mathbf{a}^{(\ell)})}{\widehat{\pi}(\ell | \bZ_i, \widehat{\mathbf{S}}^{(0)}_i)}, \quad \forall\, \ell,
\]
where the indicator function \( \mathbf{1}(\bA_i = \mathbf{a}^{(\ell)}) \) ensures that weights are only applied to observed exposure-outcome pairs. Since inverse weighting can become unstable in settings with many treatment categories, calibration weighting may be employed as a more stable alternative \citep{yang2018propensity, gao2023soft}.

\subsubsection{Step 3: Weighted Tensor Completion}

In Step 1, the initial spatial component $\widehat{\bS}^{(0)}$ is learned from an \emph{unweighted} tensor fit that does not yet account for the exposure mechanism. Using the weights from Step 2, we re-estimate both the spatial component and the tensor to correct for exposure-induced selection, yielding refined $\widehat{\bS}$ and $\widehat{\mathcal{Y}}$. Specifically, we refit the tensor with a weighted loss,
\[
\argmin_{\,\mathcal{G}, \textbf{U}_2, \textbf{U}_3, \bbb\eta_Z, \bbb\beta} \bigg\{ 
\| \sqrt{\widehat{\mathcal{W}}^{(0)}} \circ \mathcal{A} \circ (\mathcal{Y} - \mathcal{G} \times_1 (\bZ\,\bbb\eta_Z + \bbb\Phi_{1:k}\,\bbb\beta) \times_2 \bU_2 \times_3 \bU_3)\|^2_F
\bigg\},
\]
using the S-PGD. This reweighted tensor fit provides the completed potential-outcome tensor $\widehat{\mathcal{Y}}$, which can directly feed into the average treatment effector estimators with standard errors described in Section~\ref{s:ate}. 

% \textcolor{red}{iterate Steps 2 and 3? otherwise, one natural question is that it relies on a possibly biased estimator of S in weight estimation. Even using an iterated procedure, a further question is whether the algorithm converges to the truth? To mitigate such concerns, could we say in step one, we provide some initialization of S or $\phi$? e.g., step 2 can proceed with phi with a large k, right? Then one can go to step 3 $->$ step 2 again (to refine weight estimation) $->$ step 3 $->$ stop. We can stop the algorithm for one round iteration because it uses a correct initialization of $\phi$.}

\subsubsection{Spatial Projected Gradient Descent Algorithm}\label{s:comp_pgd}

Classical PGD implicitly assumes a fixed projection space for factors in any mode. In contrast, in the 
S-PGD algorithm--which we use in Steps 1 and 3 above--updates the projection space by selecting Laplacian eigenvectors in a stepwise fashion. At each step, we expand the projection by one candidate eigenvector, using warm-start from the previous solution, and retain the expansion only if it improves a BIC-type criterion. This allows the algorithm to adaptively capture spatial structure rather than committing to a pre-specified set of eigenvectors. In the above, the Tucker ranks \(r_1, r_2, r_3\) are assumed to be known. In practice, they are selected via cross-validation in Step 1 of the estimating proceduce. Complete algorithmic details, including BIC criterion, warm-start initialization scheme and stopping criteria, are provided in \textbf{Supplement B}.

\section{Theoretical Properties}\label{s:theory}

\subsection{Tensor Structural Assumptions}
Before presenting our main theorem, we introduce additional structural assumptions on the latent spatial variables and on the low-rank tensor model. These assumptions provide the technical foundation needed to establish the asymptotic bounds for our weighted spatial tensor completion estimator. 

\begin{assumption}[Unmeasured Smooth Spatial Variables]
\label{smooth_spatial}
Let $\bS \in \mathbb{R}^{N \times r_1}$ denote $r_1$ smooth graph signals over a connected, undirected graph $G = (V, E)$ with $|V| = N$. Let $\bQ$ be the normalized graph Laplacian with eigendecomposition $\bQ = \bbb\Phi \bbb\Lambda \bbb\Phi^\top$, where $\bbb\Phi = [\bbb\phi_1, \dots, \bbb\phi_N]$, where $\bbb\Phi$ has orthonormal columns, and $0 = \lambda_1 \le \lambda_2 \le \dots \le \lambda_N$. Assume that each column $\bS^{(\ell)}$ of $\bS$ admits the spectral expansion: 
\(
\bS^{(\ell)} = \sum_{j=1}^N c_j^{(\ell)} \phi_j, \quad \text{with } |c_j^{(\ell)}| \le C \cdot j^{-\alpha/\beta},
\)
for some constants $C > 0$ and $\alpha > \beta > 0$, uniformly over $\ell = 1, \dots, r_1$. 
\end{assumption}  

Assumption~\ref{smooth_spatial} formalizes the idea that unmeasured spatial confounders vary smoothly over geography. By requiring that their eigen-expansion coefficients decay at a polynomial rate, this assumption ensures that most of the spatial variation is captured by a small number of the leading eigenvectors of the Laplacian. 

\begin{assumption}[Incoherent Tensor Parameter]
\label{assume:incoherent}
Assume $\bU_1, \bU_2$ and $\bU_3$ have orthogonal columns, and there exists some constant $\mu_0$ and $\mu_1$ such that $\mathcal{Y}^* \in \mathcal{C}(r_1,r_2,r_3,\mu_0,\mu_1)$, where 
\begin{align*}
&\max \left\{ 
\frac{N}{r_1} \|\bU_1\|_{2,\infty}^2, 
\frac{L}{r_2} \|\bU_2\|_{2,\infty}^2, 
\frac{O}{r_3} \|\bU_3\|_{2,\infty}^2 \right\} \leq \mu_0, 
&\max_{k\in\{1,2,3\}} \| \mathcal{M}_{(k)}(\mathcal{G}) \| \leq \mu_1 \sqrt{\frac{NLO}{\mu_0^{3/2}(r_1r_2r_3)^{1/2}}},
\end{align*}
where for matrix $\bV$, define $\|\bV\|_{2,\infty}^2 = \max_i \|\bV_{i:}\|^2$ and $\bV_{i:}$ is the $i$-th row vector of $\bV$. 
\label{incoherent}
\end{assumption}
Assumption~\ref{assume:incoherent} is common in the matrix/tensor completion literature \citep{candes2010matrix, candes2009exact}. Incoherent matrices represent a class where each entry contains a comparable amount of information, making it feasible to complete the matrix from a small number of observations. In practice, this assumption states that information about outcomes is reasonably distributed across geographic units, exposures, and diseases.

\subsection{Asymptotic Bounds}

% \textcolor{red}{
% In order to establish the theoretical guarantees, we rely on three assumptions: (Smoothness of Spatial Variables) we assume the existence of latent spatial signals that can be represented as smooth functions over the graph Laplacian spectrum (see Assumption~\ref{smooth_spatial_app} in Appendix~\ref{appendix:assumptions}); (Latent Positivity) Each unit has a nonzero probability of being assigned to any exposure, ensuring overlap across exposure groups (see Assumption~\ref{positivity_app} in Appendix~\ref{appendix:assumptions}) (Tensor Incoherence) The true parameter tensor satisfies standard incoherence conditions widely used in the matrix/tensor completion literature \cite{candes2010matrix,candes2009exact} (see Assumption~\ref{incoherent_app} in Appendix~\ref{appendix:assumptions}).
% }

In this section, we establish the asymptotic consistency of the estimated parameter tensor $\widehat{\mathcal{Y}}$ obtained by the proposed spatial tensor completion method. We present a simplified version of our main result here; the complete theorem statement with full technical conditions is provided in \textbf{Supplement A.2}.

\begin{theorem}[Frobenius Error Bound - Simplified Version]
\label{thm:frobenius_bound} Under Assumptions~(\ref{sutva})--(\ref{assume:incoherent}), with probability at least $1 - 4\,(N+L+O)^{-2}$, our estimator satisfies: 
\begin{align}
\frac{\|\Delta^{}\|_F^2 }{NO} 
&\leq 
\max\bigg\{ A_1, A_2, A_3, A_4, A_5\bigg\}, \nonumber
\end{align}
where 
\begin{align}
&A_1 = 
\frac{\mu_1^2}{p_{\min}} \frac{\log(N + L + O)}{N}, \nonumber\\ 
&A_2 = 
C\frac{(r_1 \wedge r_2)^{1/2} r_3^{3/4}}{(r_1r_2)^{1/4}} \frac{ p_{\max} }{p_{\min}^2} \,\sigma \sqrt{r_1} k^{-\tau_2} \,\sqrt{\frac{L\log(N+L+O)}{O} }, \nonumber\\
&A_3 = 
C'\bigg(\frac{(r_1 \wedge k) r_2 r_3}{\max\{r_1 \wedge k, r_2, r_3\}} + \frac{\mu_1^2(r_1 \wedge r_2)r_3^{3/2}}{\mu_0^{3/2}(r_1r_2)^{1/2}p_{\min}^2} \bigg) \frac{ p_{\max}^2 }{p_{\min}^4} \sigma^2 \frac{\log(N+L+O)}{O}, \nonumber\\
&A_4 = 
C''\frac{p_{\max} }{p_{\min}^3} \sigma^2  \sqrt{r_1} k^{-\tau_1} \left(1 + C_5 \sqrt{\frac{\log(N+L+O)}{NO}}\right), \nonumber\\
&A_5 = 
C'''\frac{(r_1 \wedge k) r_2 r_3 (r_1 \vee r_2 \vee r_3)^2}{\max\{r_1 \wedge k, r_2, r_3\}}  \frac{1}{p_{\min}^2} \frac{(N \vee O) \log(N + L + O)}{NO}, \nonumber
\end{align}
where \( a \wedge b \), \( a \vee b \) denote the minimum and maximum of \( a \) and \( b \) respectively. 
\end{theorem}
We provide detailed interpretations of each error component in \textbf{Supplement A.2}.

\subsection{Average Treatment Effect}\label{s:ate}

After tensor completion, we estimate the average treatment effect $\theta_{\ell o}^* = \mathbb{E}[N^{-1}\sum_{i=1}^{N}(\mathcal{Y}_{i\ell o} - \mathcal{Y}_{iLo})]$ comparing exposure $\ell$ to reference exposure $L$ for outcome $o$. Given the imputed potential outcomes, the outcome imputation (OI) estimator is given by:
$$
\widehat{\theta}_{\ell o}^{\text{(OI)}} = \frac{1}{N} \sum_{i=1}^{N} ( \widehat{\mathcal{Y}}_{i\ell o} - \widehat{\mathcal{Y}}_{iLo} ).
$$
By Cauchy-Schwarz inequality and Theorem~\ref{thm:frobenius_bound}, this estimator satisfies $|\widehat{\theta}_{\ell o}^{\text{(OI)}} - \theta_{\ell o}^*| \leq \|\widehat{\mathcal{Y}} - \mathcal{Y}^*\|_F / \sqrt{N}$. However, the OI estimator does not achieve $\sqrt{N}$-consistency because tensor completion converges at a slower rate than $\sqrt{N}$ determined by the terms in Theorem~\ref{thm:frobenius_bound}. To achieve $\sqrt{N}$-consistency and enable valid asymptotic inference, we construct the augmented inverse probability weighting (AIPW) estimator \citep{bang2005doubly}:
\begin{align}
\widehat{\theta}_{\ell o}^{\text{(AIPW)}} = \frac{1}{N} \sum_{i=1}^{N} \bigg[ &\bigg( \widehat{\mathcal{Y}}_{i\ell o} + \frac{\mathbf{1}(\bA_i = \ell)}{\widehat{\pi}(\ell \mid \bZ_i, \widehat{\bS}_i)} \big(Y_{io}^{(\textbf{A}_i)} - \widehat{\mathcal{Y}}_{i\ell o}\big) \bigg) \nonumber\\
&\quad - \bigg( \widehat{\mathcal{Y}}_{iLo} + \frac{\mathbf{1}(\bA_i = L)}{\widehat{\pi}(L \mid \bZ_i, \widehat{\bS}_i)} \big(Y_{io}^{(\textbf{A}_i)} - \widehat{\mathcal{Y}}_{iLo}\big) \bigg) \bigg]. \nonumber
\end{align}
% The AIPW estimator enjoys rate double robustness: it achieves $\sqrt{N}$-consistency when either the outcome model or the propensity score model are correctly specified \citep{robins1994estimation, tsiatis2006semiparametric}. 
The AIPW estimator enjoys rate double robustness: it achieves $\sqrt{N}$-consistency when the outcome model and the propensity score model are consistently estimated, with the product of their estimation errors being of smaller order than $N^{-1/2}$, 
which permits the use of flexible modeling approaches \citep{chernozhukov2018double}. In our setting where both models share the spatial component, double robustness applies conditionally on adequate spatial approximation $\widehat{\bS}$ and depends on which term in Theorem~\ref{thm:frobenius_bound} dominates. Detailed conditions for $\sqrt{N}$-consistency % including regimes where double robustness may fail, 
are provided in \textbf{Supplement A.3}. Variance is estimated via the empirical influence function, and $100(1-\alpha)\%$ confidence intervals are constructed as $\widehat{\theta}_{\ell o}^{\text{(AIPW)}} \pm z_{\alpha/2} \sqrt{\widehat{V}_{\ell o} / N}$, where $\widehat{V}_{\ell o} = N^{-1} \sum_{i=1}^N \widehat{\text{IF}}_i^2$ with details in \textbf{Supplement A.3}. % \textcolor{red}{del \citep{tsiatis2006semiparametric, kennedy2016semiparametric}, cite the double machine learning paper here \cite{chernozhukov2018double}}

\subsection{Simulation Summary}\label{s:sim}

We conducted a simulation study to evaluate finite-sample performance; full details are provided in Supplement D. Using a $20\times 20$ spatial grid with two binary exposures and ten outcomes, we compared our proposed spatial tensor method against alternatives including non-spatial tensor completion, regression, and kriging-based method. Across settings varying in tensor complexity, strength of spatial confounding, outcome noise, and exposure overlap, our method achieved the lowest or near-lowest mean squared error while maintaining appropriate coverage for confidence intervals. These results confirm that jointly modeling exposures and outcomes within a low-rank tensor framework, combined with spectral adjustment for spatial confounding, yields efficiency gains over outcome-by-outcome approaches.

\section{PFAS Health Effect Analysis}\label{s:app}

\subsection{Study Design} 

We apply our method to examine the effects of PFAS exposures on adult chronic disease outcomes, focusing on PFOA and PFOS. Outcomes include disease rates of 13 common chronic conditions, each preprocessed by shifting to ensure positivity, applying a logarithmic transformation, and scaling the entire tensor to have zero mean and unit variance. The same transformations are applied to all competing methods. 

Our study provides broader population coverage than most prior PFAS health investigations, encompassing 5,495 public water systems serving approximately 24.8 million people across 47 states. Importantly, the exposure levels we analyze are relatively low (mean concentrations of 0.00373 ng/L for PFOA and 0.00498 ng/L for PFOS among exposed systems), reflecting typical contemporary U.S. drinking-water conditions rather than highly contaminated sites. This contrasts with earlier studies focused on heavily polluted regions with substantially higher exposures. For example, \cite{biggeri2024all} examined a population exposed to markedly elevated drinking-water PFAS concentrations in the Veneto region of Italy, while  \cite{barry2013perfluorooctanoic} investigated the Mid-Ohio Valley cohort using directly measured serum PFAS concentrations, which provide more precise exposure quantification than environmental water measurements. Our contribution lies in conducting a large-scale assessment of low-level, population-representative drinking-water exposure. However, this approach necessarily relies on environmental monitoring data rather than individual biomarker measurements, which limits our ability to capture variation in individual absorption, metabolism, and cumulative exposure from non-water sources.

\begin{figure}[tbp]
\centering
\includegraphics[width=.7\linewidth]{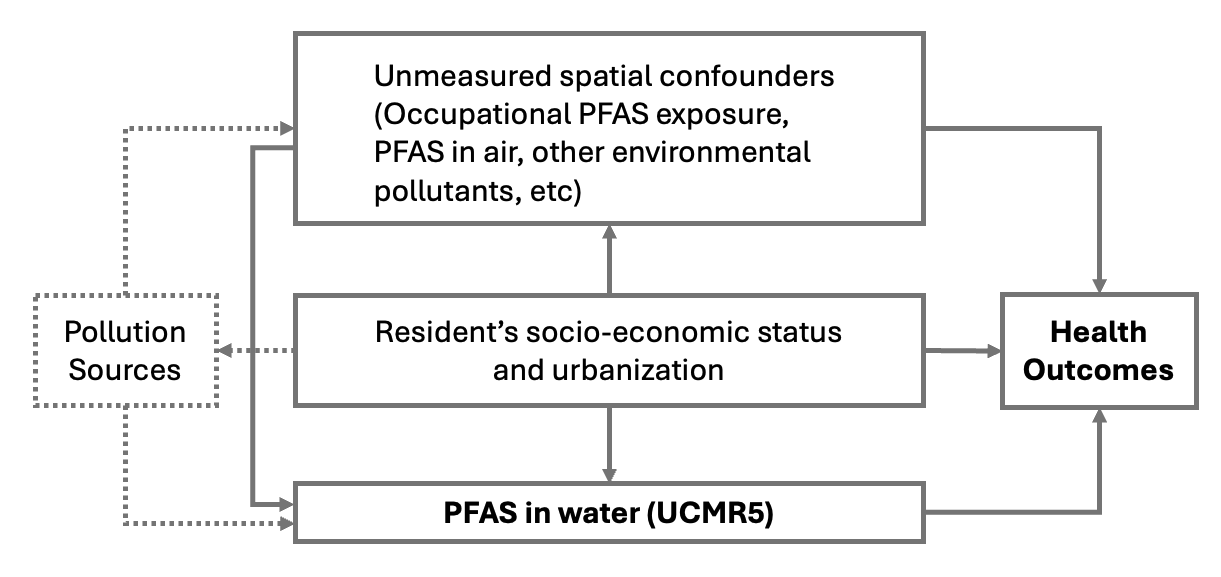}
\caption{Conceptual diagram of PFAS health effect analysis.}
\label{fig:dag}
\end{figure}

Figure \ref{fig:dag} depicts the conceptual structure of the study. Our primary focus is PFAS exposure through drinking water, as captured by `UCMR5', which is affected by both observable socio-economic and urbanization factors as well as unmeasured spatial confounders. These factors may directly influence health outcomes, introducing potential confounding. Although our analysis focuses on waterborne PFAS, the diagram also marks other exposure routes--such as occupational sources, PFAS in air, and additional pollutants--that are unobserved and approximated through their assumed spatial structure. The diagram situates our target pathway within this broader causal context and underscores the role of both measured and unmeasured determinants in shaping associations between PFAS exposure and chronic disease outcomes. 

Our final data tensor had dimensions \((5{,}495, 4, 13)\), corresponding to $N=5{,}495$ geographic units, $L=4$ exposure combinations (binary indicators for PFOA and PFOS), and $O=13$ outcomes. Spatial distance was defined using regional centroids, with adjacency constructed via a $k=4$ nearest-neighbors algorithm symmetrized by union of directed edges. The average number of neighbors per region is therefore slightly greater than 4. In \textbf{Supplement C.2} visualizes the PFAS distribution and disease distribution. 
% \textcolor{red}{Should we move these to the main text as this is a case study paper. It might be okay to put these in Section 2. You still have many pages toward the 35 page limit.}

In both the propensity score model for chemical exposure and the outcome tensor model, we incorporate covariates related to industry employment (percentage of population employed in agriculture, construction, manufacturing, wholesale, retail,  transportation, information, finance, professional, education, arts, other services, public administration) as well as a range of socio-economic indicators, including: median home value, the percentage of Hispanic/Latino residents, the percentage of foreign-born residents, median income, the percentage of individuals aged 19 to 64 with health insurance, and the percentage of residents holding a bachelor's degree or higher. Population density is included as a proxy for urbanization. The Tucker ranks are tuned via 5-fold cross-validation over the grids \(r_1 \in \{1,\dots,10\}\), \(r_2 \in \{1,\dots,3\}\), and \(r_3 \in \{1,\dots,5\}\). To improve the robustness of missingness imputation in the tensor, we implement a 5-fold cross-fitting procedure.

\subsection{Results}\label{s:app:results}

\subsubsection{Competing Methods}
% We model with four approaches: (i) augmented spatial-tensor (`Spatial-Tensor'), (ii) augment non-spatial tensor (`Tensor'), (iii) augmented spatial propensity score weighting (`Spatial-PS'), (iv) standard regression (`Standard Regression'), and (v) augmented regression with inverse probability weighting (`Augmented Regression'). \textcolor{red}{TODO: Need to be more clear what these methods are. How about a 2–3 sentence explanation of each including an equation or two in the supplement?}

We compare five estimation approaches, each combining an outcome model with a propensity score model under the augmented inverse probability weighting (AIPW) framework as described in Section \ref{s:ate}. \textbf{Spatial-Tensor}: Our proposed method using spatial tensor completion for outcomes and eigenvector-adjusted multinomial logit for propensities. \textbf{Tensor}: Tensor completion without spatial adjustment for outcomes, with standard multinomial logit for propensities. \textbf{Spatial-PS}: Standard regression fit separately per outcome, with eigenvector-adjusted multinomial logit for propensities \citep{davis2019addressing}. \textbf{Standard Regression}: Separate regressions per outcome with standard multinomial logit for propensities. \textbf{Augmented Regression}: Separate regressions per outcome, with standard multinomial logit for propensities. 
% All AIPW estimators as described in Section \ref{s:ate}. 

\subsubsection{Full Factorial Exposure Effects}

Figure~\ref{fig:result_effect} presents estimated odds ratios across the full factorial exposure structure, comparing our proposed spatial tensor method with four alternative approaches. Regression-based methods produced inflated and often implausible associations, with odds ratios exceeding 3.0 for several outcomes. In contrast, tensor-based methods provided more conservative and internally consistent estimates, with odds ratios typically below 1.05. 

The stark differences between methods reflect distinct approaches to handling missing potential outcomes and confounding. Regression-based estimate each outcome independently, neither borrowing strength across the 13 disease endpoints nor imposing structure on the counterfactual outcome surface. Tensor-based approaches, by contrast, exploit low-rank structure to pool information across outcomes and exposures simultaneously, and regularize estimates through the assumption that a small number of latent factors explain most outcome variation, leading to improved efficiency. The spatial tensor method further incorporates graph Laplacian eigenvectors to approximate unmeasured spatial confounders, removing spurious geographic associations while preserving efficiency (substantially narrower confidence intervals) gains from joint modeling.
 
The non-spatial tensor identified significant adverse associations for PFOS with four outcomes and one for PFOA, while combined PFOA+PFOS was associated with elevated odds for nearly all outcomes. Interestingly, PFOA appeared protective for asthma, whereas PFOS had the opposite effect. After adjusting for unmeasured spatial confounders via 10 selected eigenvectors, 
% \textcolor{red}{(specifically $\mathcal{F} = \{16, 18, 19, 21, 23, 24, 25, 26, 31, 89\}$; note that eigenvector 89 captures isolated regions of Colorado State)} 
the spatial tensor attenuated most associations: only two adverse health effects for PFOS (tooth loss, obesity) and one protective effect for PFOA+PFOS remained significant (COPD).

Sensitivity analyses in \textbf{Supplement C.2} demonstrate that including additional eigenvectors progressively attenuates effect estimates, indicating that unadjusted models conflate geographic confounding with causal effects. Cross-validation confirms the spatial tensor achieves superior out-of-sample prediction, validating improvements in both confounding control and predictive accuracy. 

\subsubsection{Marginal Exposure Effects}

\begin{figure}
\centering
\includegraphics[width=.8\linewidth]{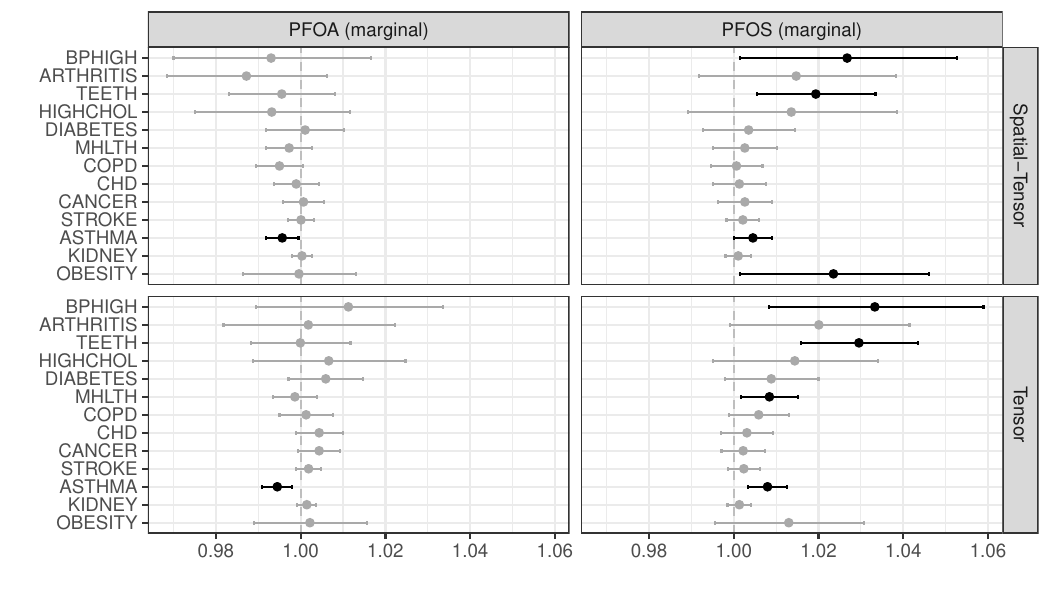}
\caption{Estimated marginal odds ratios for 13 disease outcomes under PFOA and PFOS exposures across different models. Black indicates statistically significant associations; gray indicates non-significant results.}
\label{fig:marginal}
\end{figure}

To summarize exposure effects across the factorial structure, we define marginal effects that average over the distribution of the other exposure--quantifying the expected impact of each PFAS chemical regardless of whether the other is present or absent, and relevant for regulatory policy. Figure~\ref{fig:marginal} displays the resulting marginal odds ratios with 95\% confidence intervals, comparing the spatial-tensor and non-spatial tensor models. Detailed calculations are provided in \textbf{Supplement C.2}. After adjusting for unmeasured spatial confounders via the spatial-tensor approach, PFOA showed no significant associations with any outcome except for a modest protective effect on asthma (OR = 0.995, 95\% CI: 0.991-0.999). In contrast, PFOS retained significant adverse associations with hypertension (OR = 1.026, 95\% CI: 1.001-1.052), tooth loss (OR = 1.019, 95\% CI: 1.005-1.033), asthma (OR = 1.007, 95\% CI: 1.003-1.012), and obesity (OR = 1.023, 95\% CI: 1.001-1.046). 

% \textcolor{red}{
% (Add some interpretation of the results?) 
% The comparison across methods highlights the central role of spatial confounding in large-scale environmental health studies. Non-spatial regression approaches tend to conflate geographic variation in disease prevalence with exposure effects, leading to inflated and inconsistent estimates. Joint modeling via tensor completion improves efficiency but does not, on its own, address confounding driven by unmeasured spatial factors.\\
% By explicitly incorporating spatial structure, the proposed method yields estimates that are both more conservative and more internally consistent across outcomes. While attenuation of effects may appear discouraging from a signal-detection perspective, it reflects improved confounding control and aligns with the study’s focus on low-level, population-wide exposure rather than extreme contamination scenarios.
% }

\subsubsection{Latent Factor Structure}
\begin{figure}[tbp]
\centering
\includegraphics[width=.8\linewidth]{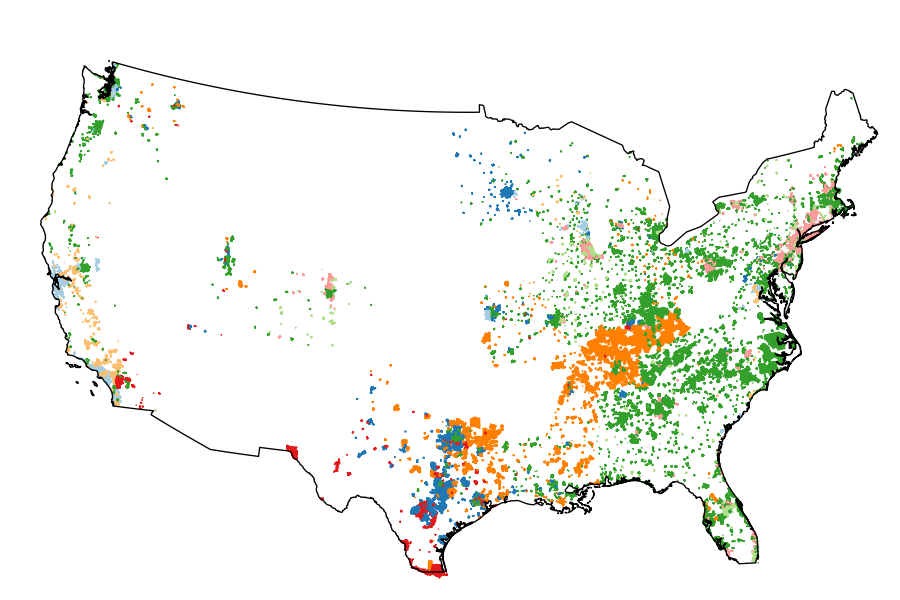}
\includegraphics[width=.8\linewidth]{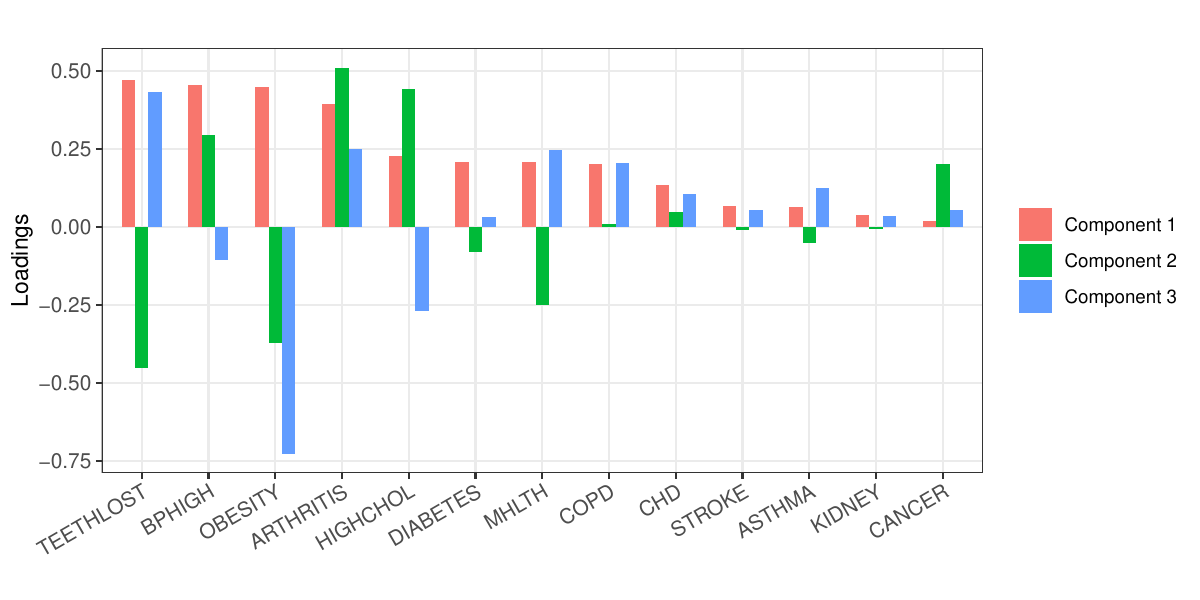}
\caption{Latent factors in the spatial tensor model. Top: Clustering of geographic units based on the location factor $\widehat\bU_1$. Bottom: Loadings in outcome factor $\widehat\bU_3$.} % Left: Two clusters from the spatial tensor model. Right: Eight clusters from the spatial tensor model.
\label{fig:result_u1}
\end{figure}

Figure~\ref{fig:result_u1} (top panel) visualizes the latent geographic structure captured by the spatial tensor model through the factor matrix $\bU_1$. We applied K-means clustering to the factor loadings to identify eight (the selected rank of $\bU_1$) distinct spatial patterns. The dominant pattern (dark green) is concentrated throughout much of the Eastern United States, and the second cluster (orange) appears concentrated in parts of the South-Central region, while smaller clusters capture localized geographic features such as the Pacific Coast, Upper Midwest. 

Figure~\ref{fig:result_u1} (bottom panel) displays the latent disease structure encoded in factor matrix $\bU_3$, which captures shared patterns of outcome variation across the 13 chronic diseases. Component 1 (red) reflects a dominant shared morbidity factor, with nearly all diseases loading positively, representing common risk pathways that affects multiple chronic conditions simultaneously. Components 2 and 3 capture more nuanced contrasts across disease groups. Interpreting these finer structural patterns in terms of specific biological mechanisms remains an important direction for future research, as they may reveal how chemical exposures align with distinct pathways of chronic disease etiology. These factors demonstrate the tensor model's ability to capture and exploit outcome structure when estimating exposure effects.

\begin{figure}[tbp]
\centering
\includegraphics[width=\linewidth]{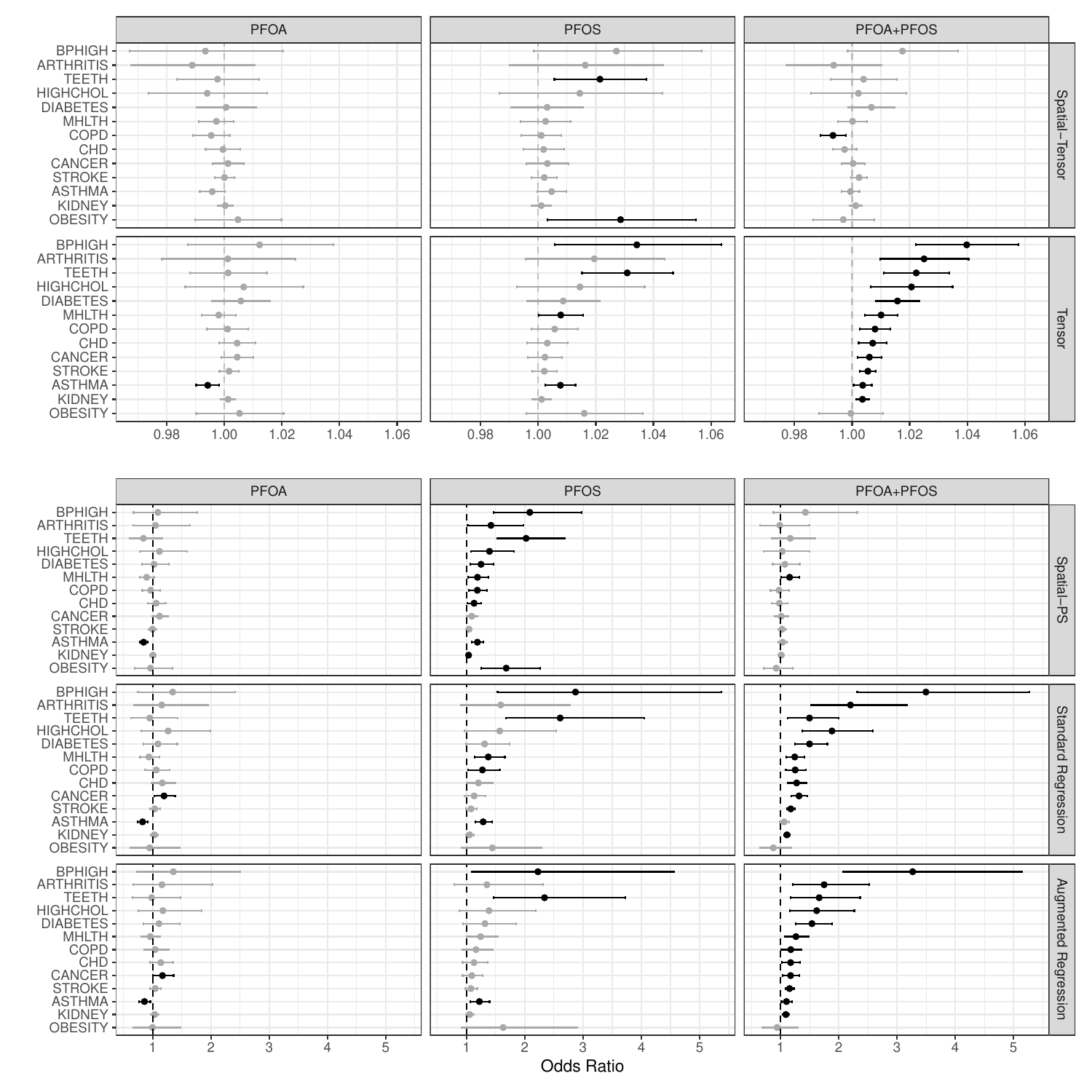}
\caption{Estimated odds ratios for 13 disease outcomes under PFOA, PFOS, and combined exposures across different models. Black indicates statistically significant associations; gray indicates non-significant results. Note that the x-axis scales differ across panels, so effect sizes are not directly comparable between the top and bottom rows.}
\label{fig:result_effect}
\end{figure}

% the cross-validation pattern \textcolor{red}{(Brian: I'd move these plots to the main text if this is going to an applied journal)} confirmed that higher ranks were needed to capture the tensor complexity; the sensitivity checks showed that including more spatial eigenvectors generally attenuated statistical significance, though estimated effects for PFOS and combined PFOA+PFOS exposures remained predominantly positive.

% Taken together, these findings suggest that unmeasured spatial confounding may inflate associations between PFAS exposures and chronic diseases, and that joint tensor modeling provides more conservative and credible effect estimates. 

% \textcolor{red}{Brian: I think for an applied journal we would maybe reduce the amount of text on the sim study and increase the amount on the read data analysis. Xiaodan: agreed, though I am not sure which figure we would include. Let's also talk about this in our meting.}

% \textcolor{red}{the spatial-ps method selects 71 eigenvectors: 0, 1, 2, 5, 6, 7, 8, 9, 10, 12, 13, 14, 15, 18, 19, 20, 21, 22, 23, 24, 25, 26, 27, 28, 29, 32, 33, 34, 35, 36, 37, 38, 39, 41, 42, 43, 44, 45, 46, 47, 54, 55, 56, 57, 58, 59, 60, 61, 62, 63, 64, 66, 68, 70, 74, 76, 77, 79, 80, 82, 83, 84, 86, 87, 89, 91, 92, 93, 96, 97, 98.}

\section{Discussion}\label{s:discussion}

To estimate the effects of PFAS chemicals on a range of health outcomes, we propose a spatial causal tensor completion framework for multiple binary exposures and outcomes, integrating a low-rank tensor structure to pool information and spectral adjustment to approximate unmeasured spatial confounders, embedded within a projected-gradient descent algorithm. This design enables causal inference in the presence of spatial confounding and pervasive missingness. Theoretically, we establish the Frobenius-error bounds for the completed outcome tensor, showing estimation error depends on spatial approximation error, propensity-score estimation error, tensor rank, and others. Extensive simulations corroborate these findings, showing our estimator achieves efficiency gains against competing approaches. 

Applied to national PFAS monitoring data linking PFOA/PFOS exposures with 13 chronic disease outcomes, our method yields substantially more conservative odds ratio estimates compared to existing alternatives. The substantial attenuation following spatial adjustment demonstrates the importance of accounting for latent geographic confounding. The alternatives identified multiple significant adverse associations (bottom panel of Figure~\ref{fig:result_effect}). After incorporating tensor structure and latent spatial confounding, our model (top panel) yielded a far sparser set of significant and attenuated effects. This systematic shift from near-universal associations to selective findings suggests that conventional approaches may substantially inflate effect estimates when spatially-structured confounders remain uncontrolled.

% \textcolor{red}{
% Our findings have implications for interpreting epidemiologic evidence more broadly. Our spatially-adjusted results for PFOS and hypertension align with consistent meta-analytic evidence linking these variables \citep{xiao2023association}. However, for many other outcomes where the epidemiologic literature remains inconsistent \citep{gui2023association, sun2018plasma, frangione2024exposure}, our results suggest inadequate control for geographic confounding may contribute to heterogeneity across studies. This highlights a key advantage of tensor-based approaches: by jointly modeling multiple outcomes and pooling information across related endpoints, our method can distinguish genuine exposure effects from spurious associations induced by shared spatial confounding.
% }

Our findings have implications for interpreting the broader PFAS epidemiologic literature. Recent meta-analyses have demonstrated fairly consistent adverse associations between PFOA/PFOS exposure and hypertension \citep{xiao2023association}, blood lipid levels \citep{liu2023associations}, and particular types of cancer \citep{bartell2021critical}. Our spatially-adjusted results for PFOS and hypertension are consistent with this meta-analytic evidence. However, for many other health outcomes, the epidemiologic evidence remains inconsistent, as exemplified by studies of diabetes \citep{gui2023association, sun2018plasma} and obesity \citep{frangione2024exposure}. Our results suggest that inadequate control for geographic confounding may have inflated effect estimates. The heterogeneity across studies may thus reflect not only differences in exposure levels, population characteristics, but also different spatial confounding bias.

Our analysis is subject to several limitations. First, disease outcomes are derived from model-based prevalence estimates rather than clinical measurements \citep{centers2023places}, introducing measurement error that likely attenuates observed associations. Second, exposure assessment relies on public water system monitoring and does not capture individual-level exposure variation or non-water exposure routes. Third, the cross-sectional linkage of 2023-2024 water monitoring data with 2017-2018 health outcomes limits causal inference, though the long environmental persistence and biological half-lives of PFAS partially mitigate temporal misalignment. Fourth, while our method adjusts for smooth spatial confounding through Laplacian eigenvectors, it cannot address all forms of unmeasured confounding, particularly individual-level factors uncorrelated with geography. Finally, our factorial design treats exposure as binary indicators rather than modeling dose-response relationships at varying concentration levels.

Several promising directions exist for extending this framework. First, incorporating continuous exposures through functional tensor representations would enable modeling of dose-response relationships rather than binary exposure indicators. Second, while the tensor framework naturally produces unit-level treatment effect estimates, developing valid inference procedures for heterogeneous and individualized causal effects remains an important methodological challenge. Third, extending the approach to longitudinal settings with time-varying exposures and outcomes would accommodate dynamic exposure patterns and enhance its applicability in environmental health and related domains.

\bigskip
\begin{center}
{\large\bf SUPPLEMENTARY MATERIAL}
\end{center}
\textbf{Supplement A:} Full Theorem, proof and interpretation.
\textbf{Supplement B:} Detailed algorithm.
\textbf{Supplement C:} Additional results with PFAS data analysis.
\textbf{Supplement D:} Simulation study. 
% \section*{Acknowledgments}
% % We thank the Editor and the anonymous referees for their constructive comments that improved the quality of this paper. 
% We thank Corinna Keeler and Shih-Ni Prim for insightful discussions and joint brainstorming on the causal mechanisms linking PFAS exposure and health outcomes.

% https://chem.echa.europa.eu/substance-search?searchText=PFAS
% https://cfpub.epa.gov/ecotox/search.cfm
% https://pubchem.ncbi.nlm.nih.gov/

% \bibliographystyle{chicago}
% \bibliography{refs}
% % \printbibliography  

\bibliography{bibliography.bib}

@article{barry2013perfluorooctanoic,
  title={Perfluorooctanoic acid (PFOA) exposures and incident cancers among adults living near a chemical plant},
  author={Barry, Vaughn and Winquist, Andrea and Steenland, Kyle},
  journal={Environmental health perspectives},
  volume={121},
  number={11-12},
  pages={1313--1318},
  year={2013},
  publisher={National Institute of Environmental Health Sciences}
}

@article{biggeri2024all,
  title={All-cause, cardiovascular disease and cancer mortality in the population of a large Italian area contaminated by perfluoroalkyl and polyfluoroalkyl substances (1980--2018)},
  author={Biggeri, Annibale and Stoppa, Giorgia and Facciolo, Laura and Fin, Giuliano and Mancini, Silvia and Manno, Valerio and Minelli, Giada and Zamagni, Federica and Zamboni, Michela and Catelan, Dolores and others},
  journal={Environmental Health},
  volume={23},
  number={1},
  pages={42},
  year={2024},
  publisher={Springer}
}

@book{national2022guidance,
  title={Guidance on PFAS exposure, testing, and clinical follow-up},
  author={NASEM and others},
  year={2022}
}

@article{bartell2021critical,
  title={Critical Review on PFOA, Kidney Cancer, and Testicular Cancer},
  author={Bartell, Scott M and Vieira, Ver{\'o}nica M},
  journal={Journal of the Air \& Waste Management Association},
  volume={71},
  number={6},
  pages={663--679},
  year={2021},
  publisher={Taylor \& Francis}
}

@article{sun2018plasma,
  title={Plasma Concentrations of Perfluoroalkyl Substances and Risk of Type 2 Diabetes: A Prospective Investigation Among US Women},
  author={Sun, Qi and Zong, Geng and Valvi, Damaskini and Nielsen, Flemming and Coull, Brent and Grandjean, Philippe},
  journal={Environmental health perspectives},
  volume={126},
  number={3},
  pages={037001},
  year={2018}
}

@article{frangione2024exposure,
  title={Exposure to perfluoroalkyl and polyfluoroalkyl substances and pediatric obesity: a systematic review and meta-analysis},
  author={Frangione, Brianna and Birk, Sapriya and Benzouak, Tarek and Rodriguez-Villamizar, Laura A and Karim, Fatima and Dugandzic, Rose and Villeneuve, Paul J},
  journal={International Journal of Obesity},
  volume={48},
  number={2},
  pages={131--146},
  year={2024},
  publisher={Nature Publishing Group UK London}
}

@article{gui2023association,
  title={Association Between Per- and Polyfluoroalkyl Substances Exposure and Risk of Diabetes: A Systematic Review and Meta-analysis},
  author={Gui, Si-Yu and Qiao, Jian-Chao and Xu, Ke-Xin and Li, Ze-Lian and Chen, Yue-Nan and Wu, Ke-Jia and Jiang, Zheng-Xuan and Hu, Cheng-Yang},
  journal={Journal of Exposure Science \& Environmental Epidemiology},
  volume={33},
  number={1},
  pages={40--55},
  year={2023},
  publisher={Nature Publishing Group US New York}
}

@article{liu2023associations,
  title={Associations Between Per- and Polyfluoroalkyl Substances Exposures and Blood Lipid Levels Among Adults—A Meta-analysis},
  author={Liu, Binkai and Zhu, Lu and Wang, Molin and Sun, Qi},
  journal={Environmental health perspectives},
  volume={131},
  number={5},
  pages={056001},
  year={2023}
}

@article{xiao2023association,
  title={Association Between Per- and Polyfluoroalkyl Substances and Risk of Hypertension: A Systematic Review and Meta-analysis},
  author={Xiao, Fang and An, Ziwen and Lv, Junli and Sun, Xiaoyi and Sun, Heming and Liu, Yi and Liu, Xuehui and Guo, Huicai},
  journal={Frontiers in public health},
  volume={11},
  pages={1173101},
  year={2023},
  publisher={Frontiers Media SA}
}

@article{wu2025latent,
  title={A Latent Factor Panel Approach to Spatiotemporal Causal Inference},
  author={Wu, Jiaxi and Franks, Alexander},
  journal={arXiv preprint arXiv:2509.10974},
  year={2025}
}

@article{prim2025spectral,
  title={A Spectral Confounder Adjustment for Spatial Regression with Multiple Exposures and Outcomes},
  author={Prim, Shih-Ni and Guan, Yawen and Yang, Shu and Rappold, Ana G and Hill, K Lloyd and Tsai, Wei-Lun and Keeler, Corinna and Reich, Brian J},
  journal={arXiv preprint arXiv:2506.09325},
  year={2025}
}

@article{wiecha2025two,
  title={Two-Stage Estimators for Spatial Confounding with Point-Referenced Data},
  author={Wiecha, Nate and Hoppin, Jane A and Reich, Brian J},
  journal={Biometrics},
  volume={81},
  number={3},
  pages={ujaf093},
  year={2025},
  publisher={Oxford University Press}
}

@article{mao2024mixed,
  title={Mixed Matrix Completion in Complex Survey Sampling under Heterogeneous Missingness},
  author={Mao, Xiaojun and Wang, Hengfang and Wang, Zhonglei and Yang, Shu},
  journal={Journal of Computational and Graphical Statistics},
  volume={33},
  number={4},
  pages={1320--1328},
  year={2024},
  publisher={Taylor \& Francis}
}

@article{yang2018propensity,
  title={Propensity Score Weighting for Causal Inference with Clustered Data},
  author={Yang, Shu},
  journal={Journal of Causal Inference},
  volume={6},
  number={2},
  pages={20170027},
  year={2018},
  publisher={De Gruyter}
}

@article{gao2023soft,
  title={Soft Calibration for Selection Bias Problems under Mixed-effects mMdels},
  author={Gao, Chenyin and Yang, Shu and Kim, Jae Kwang},
  journal={Biometrika},
  volume={110},
  number={4},
  pages={897--911},
  year={2023},
  publisher={Oxford University Press}
}

@article{bang2005doubly,
  title={Doubly Robust Estimation in Missing Data and Causal Inference Models},
  author={Bang, Heejung and Robins, James M},
  journal={Biometrics},
  volume={61},
  number={4},
  pages={962--973},
  year={2005},
  publisher={Oxford University Press}
}

@article{gao2025causal,
  title={Causal Inference on Sequential Treatments via Tensor Completion},
  author={Gao, Chenyin and Chen, Han and Zhang, Anru R and Yang, Shu},
  journal={arXiv preprint arXiv:2511.15866},
  year={2025}
}

@misc{chernozhukov2018double,
  title={Double/Debiased Machine Learning for Treatment and Structural Parameters},
  author={Chernozhukov, Victor and Chetverikov, Denis and Demirer, Mert and Duflo, Esther and Hansen, Christian and Newey, Whitney and Robins, James},
  year={2018},
  publisher={Oxford University Press Oxford, UK}
}

@article{keil2020quantile,
  title={A Quantile-based g-computation Approach to Addressing the Effects of Exposure Mixtures},
  author={Keil, Alexander P and Buckley, Jessie P and O’Brien, Katie M and Ferguson, Kelly K and Zhao, Shanshan and White, Alexandra J},
  journal={Environmental health perspectives},
  volume={128},
  number={4},
  pages={047004},
  year={2020}
}

@article{gilbert2021causal,
  title={A Causal Inference Framework for Spatial Confounding},
  author={Gilbert, Brian and Datta, Abhirup and Casey, Joan A and Ogburn, Elizabeth L},
  journal={arXiv preprint arXiv:2112.14946},
  year={2021}
}

@article{guan2023spectral,
  title={Spectral Adjustment for Spatial Confounding},
  author={Guan, Yawen and Page, Garritt L and Reich, Brian J and Ventrucci, Massimo and Yang, Shu},
  journal={Biometrika},
  volume={110},
  number={3},
  pages={699--719},
  year={2023},
  publisher={Oxford University Press}
}

@article{dupont2022spatial+,
  title={Spatial+: A Novel Approach to Spatial Confounding},
  author={Dupont, Emiko and Wood, Simon N and Augustin, Nicole H},
  journal={Biometrics},
  volume={78},
  number={4},
  pages={1279--1290},
  year={2022},
  publisher={Oxford University Press}
}

@article{marques2022mitigating,
  title={Mitigating Spatial Confounding by Explicitly Correlating Gaussian Random Fields},
  author={Marques, Isa and Kneib, Thomas and Klein, Nadja},
  journal={Environmetrics},
  volume={33},
  number={5},
  pages={e2727},
  year={2022},
  publisher={Wiley Online Library}
}

@article{keller2020selecting,
  title={Selecting a Scale for Spatial Confounding Adjustment},
  author={Keller, Joshua P and Szpiro, Adam A},
  journal={Journal of the Royal Statistical Society Series A: Statistics in Society},
  volume={183},
  number={3},
  pages={1121--1143},
  year={2020},
  publisher={Oxford University Press}
}

@inproceedings{osama2019inferring,
  title={Inferring Heterogeneous Causal Effects in Presence of Spatial Confounding},
  author={Osama, Muhammad and Zachariah, Dave and Sch{\"o}n, Thomas B},
  booktitle={International Conference on Machine Learning},
  pages={4942--4950},
  year={2019},
  organization={PMLR}
}

@article{thaden2018structural,
  title={Structural Equation Models for Dealing With Spatial Confounding},
  author={Thaden, Hauke and Kneib, Thomas},
  journal={The American Statistician},
  volume={72},
  number={3},
  pages={239--252},
  year={2018},
  publisher={Taylor \& Francis}
}

@article{carrico2015characterization,
  title={Characterization of Weighted Quantile Sum Regression for Highly Correlated Data in a Risk Analysis Setting},
  author={Carrico, Caroline and Gennings, Chris and Wheeler, David C and Factor-Litvak, Pam},
  journal={Journal of agricultural, biological, and environmental statistics},
  volume={20},
  pages={100--120},
  year={2015},
  publisher={Springer}
}

@article{kang2023partial,
  title={Partial Identification and Unmeasured Confounding With Multiple Treatments and Multiple Outcomes},
  author={Kang, Suyeon and Franks, Alexander and Audirac, Michelle and Braun, Danielle and Antonelli, Joseph},
  journal={arXiv preprint arXiv:2311.12252},
  year={2023}
}

@article{joubert2022powering,
  title={Powering Research Through Innovative Methods for Mixtures in Epidemiology (PRIME) Program: Novel and Expanded Statistical Methods},
  author={Joubert, Bonnie R and Kioumourtzoglou, Marianthi-Anna and Chamberlain, Toccara and Chen, Hua Yun and Gennings, Chris and Turyk, Mary E and Miranda, Marie Lynn and Webster, Thomas F and Ensor, Katherine B and Dunson, David B and others},
  journal={International Journal of Environmental Research and Public Health},
  volume={19},
  number={3},
  pages={1378},
  year={2022},
  publisher={MDPI}
}

@article{reich2021review,
  title={A Review of Spatial Causal Inference Methods for Environmental and Epidemiological Applications},
  author={Reich, Brian J and Yang, Shu and Guan, Yawen and Giffin, Andrew B and Miller, Matthew J and Rappold, Ana},
  journal={International Statistical Review},
  volume={89},
  number={3},
  pages={605--634},
  year={2021},
  publisher={Wiley Online Library}
}

@article{davis2019addressing,
  title={Addressing Geographic Confounding Through Spatial Propensity Scores: A Study of Racial Disparities in Diabetes},
  author={Davis, Melanie L and Neelon, Brian and Nietert, Paul J and Hunt, Kelly J and Burgette, Lane F and Lawson, Andrew B and Egede, Leonard E},
  journal={Statistical Methods in Medical Research},
  volume={28},
  number={3},
  pages={734--748},
  year={2019},
  publisher={SAGE Publications Sage UK: London, England}
}

@article{papadogeorgou2019adjusting,
  title={Adjusting for Unmeasured Spatial Confounding With Distance Adjusted Propensity Score Matching},
  author={Papadogeorgou, Georgia and Choirat, Christine and Zigler, Corwin M},
  journal={Biostatistics},
  volume={20},
  number={2},
  pages={256--272},
  year={2019},
  publisher={Oxford University Press}
}

@article{schnell2020mitigating,
  title={Mitigating Unobserved Spatial Confounding When Estimating the Effect of Supermarket Access on Cardiovascular Disease Deaths},
  author={Schnell, Patrick M and Papadogeorgou, Georgia},
  journal={arXiv preprint arXiv:1907.12150},
  year={2020}
}

@article{fenton2021per,
  title={Per- and Polyfluoroalkyl Substance Toxicity and Human Health Review: Current State of Knowledge and Strategies for Informing Future Research},
  author={Fenton, Suzanne E and Ducatman, Alan and Boobis, Alan and DeWitt, Jamie C and Lau, Christopher and Ng, Carla and Smith, James S and Roberts, Stephen M},
  journal={Environmental toxicology and chemistry},
  volume={40},
  number={3},
  pages={606--630},
  year={2021},
  publisher={Wiley Periodicals, Inc. Hoboken}
}

@article{gaines2023historical,
  title={Historical and Current Usage of Per- and Polyfluoroalkyl Substances (PFAS): A Literature Review},
  author={Gaines, Linda GT},
  journal={American Journal of Industrial Medicine},
  volume={66},
  number={5},
  pages={353--378},
  year={2023},
  publisher={Wiley Online Library}
}

@article{walker2021package,
  title={Package ‘tidycensus’},
  author={Walker, Kyle and Herman, Matt and Eberwein, Kris and Walker, Maintainer Kyle},
  journal={MIT},
  year={2021}
}

@article{bobb2015bayesian,
  title={Bayesian Kernel Machine Regression for Estimating the Health Effects of Multi-Pollutant Mixtures},
  author={Bobb, Jennifer F and Valeri, Linda and Claus Henn, Birgit and Christiani, David C and Wright, Robert O and Mazumdar, Maitreyi and Godleski, John J and Coull, Brent A},
  journal={Biostatistics},
  volume={16},
  number={3},
  pages={493--508},
  year={2015},
  publisher={Oxford University Press}
}

@misc{centers2023places,
  title={PLACES: Local Data for Better Health, Census Tract Data 2023 Release},
  author={{CDC}},
  year={2023},
  publisher={Centers for Disease Control and Prevention website}
}

@misc{epapfastool,
  author    = {{US EPA}},
  year      = {2024},
  title     = {EPA’s PFA Analytics Tool},
  url       = {https://echo.epa.gov/trends/pfas-tools},
  note      = {Accessed: 2024-08-12}
}

@article{zhen2024nonnegative,
  title={Nonnegative Tensor Completion for Dynamic Counterfactual Prediction on COVID-19 Pandemic},
  author={Zhen, Yaoming and Wang, Junhui},
  journal={The Annals of Applied Statistics},
  volume={18},
  number={1},
  pages={224--245},
  year={2024},
  publisher={Institute of Mathematical Statistics}
}

@article{abadie2024doubly,
  title={{Doubly Robust Inference in Causal Latent Factor Models}},
  author={Abadie, Alberto and Agarwal, Anish and Dwivedi, Raaz and Shah, Abhin},
  journal={arXiv preprint arXiv:2402.11652},
  year={2024}
}

@article{poulos2021retrospective,
  title={Retrospective Causal Inference via Matrix Completion, With an Evaluation of the Effect of European Integration on Cross-Border Employment},
  author={Poulos, Jason and Albanese, Andrea and Mercatanti, Andrea and Li, Fan},
  journal={arXiv preprint arXiv:2106.00788},
  year={2021}
}

@article{candes2010matrix,
  title={Matrix Completion With Noise},
  author={Candes, Emmanuel J and Plan, Yaniv},
  journal={Proceedings of the IEEE},
  volume={98},
  number={6},
  pages={925--936},
  year={2010},
  publisher={IEEE}
}

@article{candes2009exact,
  title={Exact Matrix Completion via Convex Optimization},
  author={Candes, Emmanuel and Recht, Benjamin},
  journal={Communications of the ACM},
  volume={55},
  number={6},
  pages={111--119},
  year={2009},
  publisher={ACM New York, NY, USA}
}

@article{agarwal2020synthetic,
  title={Synthetic Interventions},
  author={Agarwal, Anish and Shah, Devavrat and Shen, Dennis},
  journal={arXiv preprint arXiv:2006.07691},
  year={2020}
}

@article{kolda2009tensor,
  title={Tensor Decompositions and Applications},
  author={Kolda, Tamara G and Bader, Brett W},
  journal={SIAM review},
  volume={51},
  number={3},
  pages={455--500},
  year={2009},
  publisher={SIAM}
}

@article{rubin1974estimating,
  title={Estimating Causal Effects of Treatments in Randomized and Nonrandomized Studies.},
  author={Rubin, Donald B},
  journal={Journal of educational Psychology},
  volume={66},
  number={5},
  pages={688},
  year={1974},
  publisher={American Psychological Association}
}

@article{auerbach2022tensor,
  title={Tensor Completion for Causal Inference with Multivariate Longitudinal Data: A Reevaluation of COVID-19 Mandates},
  author={Auerbach, Jonathan and Slawski, Martin and Zhang, Shixue},
  journal={arXiv preprint arXiv:2203.04689},
  year={2022}
}

@article{gao2024causal,
  title={Causal Customer Churn Analysis with Low-rank Tensor Block Hazard Model},
  author={Gao, Chenyin and Zhang, Zhiming and Yang, Shu},
  journal={arXiv preprint arXiv:2405.11377},
  year={2024}
}

\end{document}